\documentstyle[emulateapj,psfig,apjfonts]{article}

\def\gs{\mathrel{\raise0.35ex\hbox{$\scriptstyle >$}\kern-0.6em
\lower0.40ex\hbox{{$\scriptstyle \sim$}}}}
\def\ls{\mathrel{\raise0.35ex\hbox{$\scriptstyle <$}\kern-0.6em
\lower0.40ex\hbox{{$\scriptstyle \sim$}}}}

\submitted{Received 2001 August 31; accepted 2001 September 24}

\lefthead{Ivison et al.}
\righthead{A high-resolution imaging study of SMM\,J14011+0252}

\begin{document}

\title{Locating the starburst in the SCUBA galaxy, SMM\,J14011+0252}

\author{R.\ J.\ Ivison\altaffilmark{1}, Ian Smail\altaffilmark{2},
D.\ T.\ Frayer\altaffilmark{3}, J.-P.\ Kneib\altaffilmark{4}
\&  A.\ W.\ Blain\altaffilmark{5}}
\altaffiltext{1}{Astronomy Technology Centre, Royal Observatory,
Blackford Hill, Edinburgh EH9 3HJ, UK}
\altaffiltext{2}{Department of Physics, University of Durham, South 
Road, Durham DH1 3LE, UK}
\altaffiltext{3}{{\em SIRTF} Science Centre, California Institute of
Technology, IPAC, MS\,220-06, Pasadena, CA 91125, USA}
\altaffiltext{4}{Observatoire Midi-Pyr\'en\'ees, UMR 5572,
14 Avenue E.\ Belin, F-31400 Toulouse, France}
\altaffiltext{5}{Astronomy 105-24, California Institute of
Technology, Pasadena, CA 91125, USA}

\setcounter{footnote}{5}

\begin{abstract}
We present new, multi-wavelength, high-resolution imaging of the
luminous, submillimeter (submm) galaxy, SMM\,J14011+0252, an
interacting starburst at $z=2.56$.  Our observations comprise optical
imaging from the {\it Hubble Space Telescope}, sensitive radio mapping
from the Very Large Array and CO observations from the Owens Valley
Radio Observatory and Berkeley-Illinois-Maryland Array.  Aided by
well-constrained gravitational amplification, we use these new data to
map the distribution of gas and both obscured and unobscured
starlight. The maps show that the gas and star formation are extended
on scales of $\gs$10\,kpc, much larger than starbursts seen in local
ultraluminous galaxies, and larger than the rest-frame UV-bright
components of SMM\,J14011+0252, J1/J2. The most vigorous star
formation is marked by peaks in both the molecular gas and radio
emission, $\sim$1$''$ north of J1/J2, in the vicinity of J1n, an
apparent faint extension of J1.  Using new sub-$0.5''$ $K$-band
imaging from UKIRT, we identify J1n as an extremely red object
(ERO). We suggest that while J1 and J2 are clearly associated with the
submm source, they are merely windows through the dust, or unobscured
companions to a large and otherwise opaque star-forming
system. Hence, their rest-frame UV properties are unlikely to be relevant
for understanding the detailed internal physics of the starburst. 
\end{abstract}


\section{Introduction}

The resolution into individual sources of the majority of the optical,
submm and X-ray backgrounds (Madau \& Pozzetti 2000; Blain et al.\
1999; Brandt et al.\ 2001) means that we can now investigate the
relationship and balance between obscured and unobscured star
formation and AGN activity at high redshifts.  For example, the
limited overlap between X-ray and submm populations has been taken as
evidence that the bulk of the submm background arises from reprocessed
starlight (Fabian et al.\ 2000; Barger et al.\ 2001; Almaini et al.\
2001). While the broad equality between the intensity of the optical
and far-IR background radiation has then been used to argue that
comparable amounts of star formation occur in obscured and unobscured
environments (Hauser et al.\ 1998).  A major unresolved issue is the
relationship between the obscured and unobscured modes of star
formation, as represented by SCUBA-selected galaxies and Lyman-break
galaxies, respectively (Smail, Ivison, \& Blain 1997; Steidel,
Pettini, \& Hamilton 1995). Is unobscured activity simply the tip of
the iceberg, and can it be used to predict the obscured component
(Adelberger \& Steidel 2000), or do Lyman-break galaxies represent a
physically distinct population with no meaningful relationship with
galaxies selected in the submm (van der Werf et al.\ 2001; Meurer et
al.\ 2001; also see Bell et al.\ 2001)?  This has important
implications for the relative contributions of optical- and
submm-selected galaxies to the history of star formation (Blain et
al.\ 1999b; Chapman et al.\ 2000; Peacock et al.\ 2001).

\setcounter{figure}{0}
%
%
\begin{figure*}[htb]
\centerline{\psfig{file=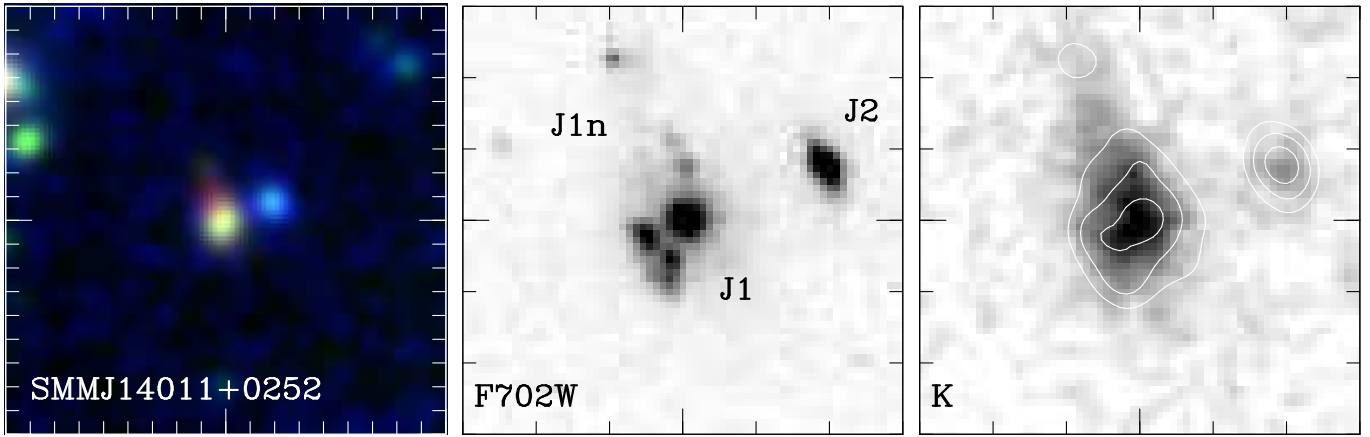,width=7.1in,angle=0}}
\vspace*{-2mm}
\caption{\footnotesize
Three views of SMM\,J14011+0252: a true-color $UR_{702}K$ image (using
$U$-band data from I00), the {\it HST} F702W frame and the UKIRT
$K$-band image.  The true-color frame is $18''\times 18''$, while the
F702W/$K$-band views are zoomed to give a $6''\times 6''$ field to
better illustrate the internal structure of this galaxy.  We identify
the two UV-bright components of SMM\,J14011+0252 on the F702W panel:
the morphologically complex J1 and the compact, blue component J2, as
well as the diffuse, very red component, J1n. We overlay a contour plot of the
seeing-matched F702W image on the $K$-band panel to contrast the
optical/IR morphologies (the contours are in 1-mag increments starting
at $\mu_R=24.0$\,mag\,arcsec$^{-2}$).  North is up; east is left;
minor tickmarks denote 1$''$ increments.}
\end{figure*}

To address this issue we need to study the properties of individual
submm galaxies. Unfortunately, progress has been constrained by their
faintness in the rest-frame UV and the resultant dearth of
spectroscopic redshifts and detailed morphologies for more than a
handful of systems (Ivison et al.\ 1998, 2000a --- I00).  For the
moment all we have to work with are a few, perhaps atypical,
optically-bright galaxies with precise redshifts. There are also
concerns about the over-representation of AGN in the current samples
of spectroscopically-identified submm galaxies, reflecting a bias from
the relative ease of measuring redshifts for their strong-lined
spectra.  Although star formation, rather than the AGN, is thought to
dominate the energetics of these systems, the presence of an AGN
clearly complicates any detailed analysis. We are therefore left with
only a single example from current surveys of the class of galaxy
which is expected to comprise the majority of the submm population:
the $z=2.56$, gas-rich, luminous starburst, SMM\,J14011+0252 (I00),
although even this appears unusually bright in the rest-frame UV.

SMM\,J14011+0252, lying behind the core of the $z=0.25$ cluster
A\,1835, has an 850-$\mu$m flux of 6\,mJy, after correcting for
amplification by the foreground cluster lens (a mean factor of
$2.5\times$, dominated by amplification along position angle
9$^{\circ}$), showing that this is an ultraluminous IR galaxy (ULIRG)
with a far-IR luminosity of $L_{\rm FIR}\sim 6\times
10^{12}$\,L$_\odot$, a star-formation rate (SFR) of $\sim
10^3$\,M$_\odot$\,yr$^{-1}$ and a large dust mass, $M_{\rm d} \sim
10^8$--$10^9$\,M$_{\odot}$.  I00 present detailed observations of this
submm-selected source, which they identified with an
interacting/merging pair of galaxies, J1 and J2, separated by $2.1''$,
with a combined absolute magnitude of $M_R =
-25.0$\footnote{Throughout we will assume $q_o=0.5$ and
$H_o=50$\,km\,s$^{-1}$\,Mpc$^{-1}$, giving a scale of $1''\equiv
7.7$\,kpc at $z=2.56$.  We quote all magnitudes and linear scales as
observed, rather than corrected for lens amplification, unless
otherwise stated.}.  J2 contributes the bulk of the luminosity in the
UV, with the redder J1 dominating the near-IR.  Optical and near-IR
spectra of J1 and J2 show no hint of AGN characteristics (I00) and the
system is undetected in hard X-ray observations with {\it Chandra}
(Fabian et al.\ 2000), supporting the contention that its luminosity
is predominantly produced by an intense starburst. The accurate
redshift for this system allowed Frayer et al.\ (1999) to detect
CO(3$\rightarrow$2) emission, confirming the presence of a large gas
reservoir, $M({\rm H}_2)\sim 10^{11}$\,M$_\odot$ (probably a lower
limit based on evidence of cold gas from CO(1$\rightarrow$0) emission
in another source, Papadopoulos et al.\ 2001).

Although the observational dataset on SMM\,J14011+0252 is the most
comprehensive available for any submm galaxy, there is still
considerable freedom in interpreting the results.  In particular, we
need to identify the location of the intense far-IR emission within
the system, and the attendant gas reservoir, to understand the exact
relationship between the UV and far-IR emission (Adelberger \& Steidel
2000) and determine if they are coming from physically disjoint
components.  Unfortunately, the published observations had
insufficient spatial resolution to answer this question.  We have
therefore undertaken a campaign to image SMM\,J14011+0252 at high
resolution in the radio, millimeter (CO) and optical/near-IR wavebands
to investigate in detail the distribution of obscured and unobscured
star formation in this galaxy.

We describe our new high-resolution images in \S2, analyse these in
\S3 and present our conclusions in \S4.

\section{Observations and Reduction}

\setcounter{figure}{1}
%
%
\begin{figure*}[htb]
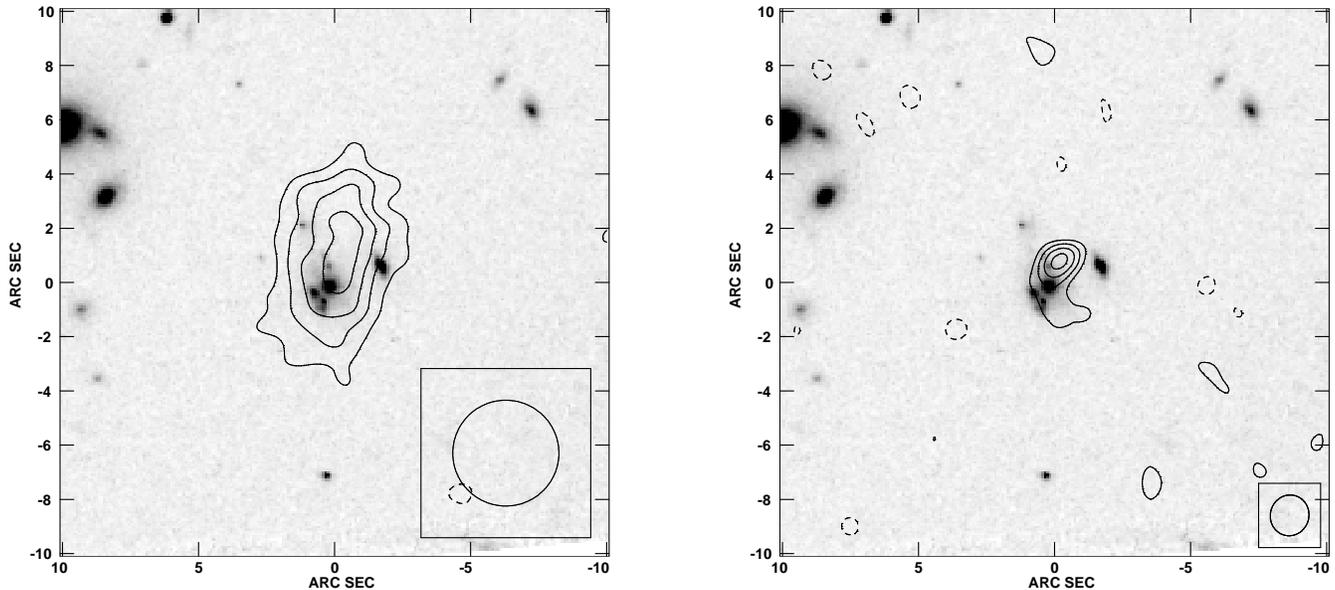

\centerline{\psfig{file=f2a.eps,width=3.2in,angle=0}
\hspace*{0.5in}
\psfig{file=f2b.eps,width=3.2in,angle=0}}
\vspace*{-3mm}
\caption{\footnotesize (a) Combined OVRO/BIMA CO image of
SMM\,J14011+0252, with a noise level of 0.8\,mJy\,beam$^{-1}$; (b) VLA
1.4-GHz map, with a noise level of 12\,$\mu$Jy\,beam$^{-1}$. The CO
and radio emission are co-spatial, peaking $\sim$1$''$ north of
J1. The CO subtends $6.6''\pm 1.4''$ in the north-south direction, the
radio somewhat less. All contours are plotted at $-3.5$, $-2.5$, 2.5,
3.5, 4.5, 5.5 $\times \sigma$; positions are relative to the position
of J1, shown as a greyscale; synthesized beams are shown.}
\end{figure*}

\subsection{Optical and Near-IR Imaging}

The {\it HST}\,\footnote{This paper is based upon observations
obtained with the NASA/ESA {\it Hubble Space Telescope} which is
operated by STScI for the Association of Universities for Research in
Astronomy, Inc., under NASA contract NAS5-26555.} imaging of
SMM\,J14011+0252, in the field of A\,1835, comprises three orbits in
the F702W filter, giving a combined integration time of 7.5\,ks.  We
adopt the Vega-based {\it WFPC2} photometric system, $R_{702}$, from
Holtzman et al.\ (1995).  The final WFPC2 frame (see Fig.~1) has an
effective resolution of 0.15$''$ and a 3-$\sigma$ detection limit
within a 2$''$--diameter photometry aperture of $R_{702}\sim26.6$.  We
also exploit new, near-IR imaging of SMM\,J14011+0252. Using UFTI on
the UKIRT,\footnote{The United Kingdom Infrared Telescope (UKIRT) is
operated by the Joint Astronomy Centre on behalf of the Particle
Physics and Astronomy Research Council.} Smith et al.\ (2001) obtained
a 6.5-ks $K$-band exposure during 2001 April 4--7.  The image
(Fig.~1c) has a 3-$\sigma$ limit of $K\sim 21.2$ and a FWHM of
0.45$''$.  The reduction and analysis of both the {\it HST} and UKIRT
data is described in full in Smith et al.\ (2001).

\subsection{Centimeter and Millimeter Mapping}

The initial Owens Valley Millimeter Array (OVRO)\footnote{OVRO is
operated by the California Institute of Technology and is supported by
NSF grant AST 9981546.} CO observations taken of J1/J2 in 1998 showed
evidence for a possible extended component in the north-south
direction (Frayer et al.\ 1999).  We subsequently obtained 78\,hr of
high-resolution CO observations at OVRO and the
Berkeley-Illinois-Maryland Array (BIMA)\footnote{The BIMA array is
operated by the Berkeley-Illinois-Maryland Association under funding
from the NSF.} to constrain the CO morphology and position of the
source (Table~1).

%
%
{\scriptsize
\begin{center}
\centerline{\sc Table 1}
\centerline{\sc Millimeter Observations of SMM\,J14011+0252}
\begin{tabular}{cccccc}
\hline\hline
\noalign{\smallskip}
Observing  &    Array    &     Baseline   &   Time on   &    RMS &  $\theta_b^a$ \cr
Dates      &             &    Lengths (m) &  Source (h) &  (mJy)&   ($''$) \cr
\noalign{\smallskip}
\hline
\noalign{\smallskip}
1998 Oct--Dec & OVRO(L+E)    & 15--119      &   41.4  &      1.0  &  $6.5\times  4.3$\cr
1999 Dec     & BIMA(A)       & 80--1730     &    16.7  &      2.0 &  $0.7\times 0.6$\cr 
2000 Feb--Mar & OVRO(H+U)    & 35--483      &    19.8 &       1.8 &   $2.4\times 1.9$\cr
\hline
\end{tabular}
\end{center}
\addtolength{\baselineskip}{-3pt}
$^a$ The rms and synthesized beam sizes ($\theta_b$) are for a
natural-weighted map averaged over the CO line width of
320\,km\,$s^{-1}$.  The data were taken in four OVRO configurations
(L, E, H, U) and the highest resolution BIMA configuration (A).

}
\medskip

Fig.~2a shows the integrated CO map of the combined OVRO and BIMA
data.  In order to resolve the galaxy while maintaining good signal to
noise, we adopt a natural-weighting scheme with a UV taper of
120\,$k\lambda$ using {\sc aips}.  We achieved an rms noise level
0.8\,mJy\,beam$^{-1}$ integrated over the CO emission-line width of
320\,km\,s$^{-1}$, with a beam of $3.9''\times 3.9''$.

The A\,1835 field was also mapped with the National Radio Astronomy
Observatory's (NRAO) Very Large Array (VLA),\footnote{NRAO is operated
by Associated Universities Inc., under a cooperative agreement with
the National Science Foundation.} during 1998 April and 2000 November.
After standard calibration and editing of the data and their
associated weights using {\sc aips}, the wide-field imaging task, {\sc
imagr}, was used to map the central $10'\times 10'$ field (see, e.g.\
Richards 2000). The resulting map (Fig~2b) has a noise level of
12\,$\mu$Jy\,beam$^{-1}$ with 1.45$''$ resolution.

\subsection{Astrometry}

To reliably compare the positions of the components of
SMM\,J14011+0252 we have revisited the astrometry from I00 to tie the
optical and near-IR images directly to the radio coordinate frame.  By
identifying bright radio sources within the {\it HST} field we align
the frames to a precision of $\sim$0.3$''$.  We then measure the
position of J1 to be $\rm 14^h 01^m 04.^{\!\!s}95 \pm 0.^{\!\!s}02,
+02^{\circ}$ 52$'$ 24.$\!\!''$0 $\pm$ 0.3$''$ (J2000).

The peak of the 1.4-GHz radio source lies at $\rm 14^h 01^m
04.^{\!\!s}92, +02^{\circ}$ 52$'$ 24.$\!\!''$8 (J2000), with an
uncertainty of $\pm 0.2''$, coincident with the CO emission which
peaks at $\rm 14^h 01^m 04.^{\!\!s}93, +02^{\circ}$ 52$'$ 25.$\!\!''$0
(J2000), with an uncertainty of $\pm 0.7''$.

\section{Analysis and Discussion}

The {\it HST} imaging of SMM\,J14011+0252, aided by the gravitational
amplification from A\,1835, provides a sub-kpc scale view of the
components of this submm source in the rest-frame UV, $\sim
2000$\,\AA.  This shows that J1 is well resolved, with a half-light
diameter of $1.34'' \pm 0.08''$ ($3.3\pm 0.2$\,kpc in the source
plane). As with another submm galaxy, Lockman 850.1 (Lutz et al.\
2001), its morphology is complex: the main component appears to be
relatively regular, with an extended envelope.  Superimposed on this
are several bright knots, in two groups: one to the south-east and the
other to the north (which also appears in $K$).  The northern group
points towards another bright, resolved blue knot, $2.5''$ away.  This
knot defines the outer limit of a low-surface-brightness extension,
J1n, to the north of J1, which has an extent of 10\,kpc in the source
plane and which is much more prominent in $K$ than $R$ (Fig.~1). We
note, as did Ivison et al.\ (1998), that red plumes such as this may
be due to line emission: probably H\,$\alpha$ at 2.34\,$\mu$m (I00),
which could be mapped in the future using near-IR integral-field
spectroscopy.

The second bright component, J2, is more luminous than J1 at
rest-frame wavelengths of $\ls 2000$\,\AA, but rapidly becomes fainter
in redder wavebands.  J2 has a compact appearance, with a half-light
diameter of $0.50'' \pm 0.06''$.  The morphology of J2 suggests it
could be an edge-on disk extending out to $\sim 0.5''$, with a
high-surface-brightness nucleus, though the prominence of this
extension may result from the higher north-south amplification.

Photometry from the seeing-matched F702W and $K$-band frames gives:
J1, $(R_{702}-K)=3.29\pm 0.03$ and J2, $(R_{702}-K)=2.12\pm 0.05$,
both in apertures twice the size of their respective half-light
diameters.  J1n is much redder than either of the bright components
with $(R_{702}-K)=5.03\pm 0.10$, very similar to the extreme colors
seen in other submm galaxies\footnote{SMM\,J14009+0252 is a 14.5-mJy
850-$\mu$m source in the same field as SMM\,J14011+0252. It has a very
faint near-IR counterpart (J5) which was, and remains, undetected at
optical wavelengths.  We can now place a 3--$\sigma$ limit on its
color of $(R_{702}-K)\geq 5.8$ in a $2''$-diameter aperture, meaning
that this galaxy is an ERO or a Class~0/I submm source (Ivison et al.\
2000b). This brings the fraction of EROs in the Smail et al.\ sample
to $\ge$27 per cent ($\ge$44 per cent for the 4$\sigma$ sample).}
(Smail et al.\ 1999).

As shown in Fig.~2, the peaks of the CO and radio emission in our
high-resolution maps lie $\sim$1$''$ north of J1 and $\sim$2$''$ east
of J2.  Given the precision of our astrometry, these offsets are
significant and so we can rule out the starburst arising from either
of these components. The radio/CO emission are not coincident with the
ERO, J1n, but are more consistent with this component than with any
other. Thus the bulk of the luminosity, gas and dust appear be located 
close to, or within, the extremely red component.

I00 estimated the SFR for $\ge8$\,M$_{\odot}$ stars in
SMM\,J14011+0252, based on far-IR and radio fluxes, of $\sim
10^3$\,M$_{\odot}$\,yr$^{-1}$. In contrast, the H$\alpha$ luminosity
indicates only $\sim 10^2$\,M$_{\odot}$\,yr$^{-1}$. They concluded
that the likely cause of this discrepency is significant dust
extinction, extending to rest-frame $\sim$6600\AA (observed $K$-band),
which obscures the most vigorously star forming regions within the
galaxy. This is consistent with the faint appearance of the
star-forming region in our $K$-band image and its complete absence in
the {\it HST} frame, as well as the conclusions of studies which
constrast the very different appearance of local starbursts in the
optical and the near-IR/radio.  We suggest that the known optical
components, J1/J2, merely represent foreground star-forming knots, or
windows through the dust in a large and otherwise opaque star-forming
complex.

Since the starburst is obscured even at $K$, we must turn to our
long-$\lambda$ observations to investigate its morphology.  The
distributions of both the radio and CO emission are extended roughly
north--south, consistent with alignment of J1--J1n, with the emission
resolved along this axis. After deconvolving the beam, the molecular
gas traced by CO subtends $6.6''\pm1.4''$ along the major axis
(Fig.~2a). Assuming conservatively that the amplification is entirely
north-south, the molecular gas must cover at least 20\,kpc. The best
Gaussian fit to the 1.4-GHz emission gives $2.3'' \times 1.5''$ ({\sc
fwhm}), although a fainter emission extends to the south (Fig.~2b). We
stress that several nearby radio sources remain unresolved in our map,
so this extension is real. This strongly suggests that the far-IR
emission is powered by a starburst: the vast molecular gas reservoir
appears to be converted into stars and ultimately into SNe on a
galaxy-wide scale. The growing evidence of highly extended starbursts
in the submm galaxy population (Ivison et al., in prep) indicates that
physical conditions (gas densities, dust shielding, wind formation) in
these systems will be very different from those expected on the basis
of local, compact ($\le 1$\,kpc) ULIRGs.

\section{Conclusions}

We have presented new high-resolution optical, CO and radio
observations of the $z=2.56$ SCUBA-selected ULIRG, SMM\,J14011+0252.
Our new $K$-band imaging has allowed us to resolve an ERO component
within this system ($R_{702}-K=5.0$). Of the known optical/near-IR
components, this heavily dust-obscured component is closest to the
starburst, as probed by our high-resolution CO and radio maps.

These observations are relevant for the issue of the relationship
between the Lyman-break and submm galaxy populations.  In this paper
we have studied a galaxy which has been claimed to represent the
transition between these two populations, with UV properties that
confirm the reliability of large extinction corrections of
the type applied to Lyman-break galaxies (Adelberger \& Steidel 2000).
Our higher resolution data have demonstrated that the properties of
SMM\,J14011+0252 are much more complex than previously thought, with
the bulk of the luminosity coming from an extremely red component
which is not seen in rest-frame UV observations, or possibly from a
region which is blank even in the near-IR. While the optically bright
components, J1 and J2, are clearly associated with the submm source,
they appear as distinct regions within the system and their rest-frame
UV properties are therefore unlikely to provide a reliable indication
of the physical processes occuring within the highly obscured
starburst. J1, J2 and even J1n may merely mark windows through the
dust in an otherwise opaque star-forming complex. This calls into
question the validity of using UV-selected objects to trace the
evolution of obscured star formation at high redshifts.

Nevertheless, in the absence of the less obscured components, J1 and
J2, this system would have been difficult to study in the detail
achieved here.  The presence of UV-bright companions may prove to be
the only practical way to measure redshifts for submm galaxies for the
foreseeable future (see the discussion of J5 in I00; cf.\ Blain et
al.\ 2000; Townsend et al.\ 2001).

New information on the luminous submm source, SMM\,J14009+0252,
previously suggested as a radio-loud AGN with ERO characteristics,
confirm that this galaxy is an ERO, raising the fraction of known
extremely red counterparts to almost half of the most robust parent
sample.

High-redshift dusty galaxies such as SMM\,J14011+0252 and
SMM\,J14009+0252 account for $\gs 50$ per cent of the submm background
(Blain et al.\ 1999), leaving room for only a modest contribution
from the submm counterparts of UV-selected systems (Peacock et al.\
2000).  This is in sharp contrast to the situation in the local
Universe, where very luminous dusty starbursts contribute a minor
fraction of the total energy output, underlining the rapid evolution
which must occur in the ULIRG population (Smail et al.\ 1997).
Identifying the physical process responsible for this rapid increase
in star formation, as well as the reason for the much larger physical
scale of high-redshift starbursts compared to local examples, will
yield important insights into the formation of massive galaxies.

\acknowledgments

We thank Leo Blitz, Harald Ebeling, Frazer Owen, Nick Scoville, Graham
Smith, Jack Welch and Mel Wright for help and advice. IRS and JPK
acknowledge support from the Royal Society, the Leverhulme Trust and
CNRS.

\end{document}